\documentclass{article}
\usepackage{spconf,graphicx}
\usepackage{amsmath}
\usepackage{multirow}
\usepackage{balance}

\usepackage{algorithmic}
\usepackage{colortbl}
\usepackage{hyperref}
\usepackage{booktabs}
\usepackage[ruled]{algorithm2e}
\usepackage{balance}
\usepackage{comment}

\usepackage{color}
\usepackage{soul}
\usepackage{subcaption}
\usepackage{multirow}
\usepackage{pifont}
\usepackage{xcolor}

\PassOptionsToPackage{hyphens}{url}\usepackage{hyperref}

\ninept

\title{MorphFader: Enabling Fine-grained Controllable Morphing with Text-to-Audio Models}
\name{Purnima Kamath, Chitralekha Gupta, Suranga Nanayakkara}

\address{National University of Singapore, Singapore}

\begin{document}
\maketitle
\begin{abstract}
Sound morphing is the process of gradually and smoothly transforming one sound into another to generate novel and perceptually hybrid sounds that simultaneously resemble both. Recently, diffusion-based text-to-audio models have produced high-quality sounds using text prompts. However, granularly controlling the semantics of the sound, which is necessary for morphing, can be challenging using text. In this paper, we propose \textit{MorphFader}, a controllable method for morphing sounds generated by disparate prompts using text-to-audio models. By intercepting and interpolating the components of the cross-attention layers within the diffusion process, we can create smooth morphs between sounds generated by different text prompts. Using both objective metrics and perceptual listening tests, we demonstrate the ability of our method to granularly control the semantics in the sound and generate smooth morphs.

\end{abstract}
        
\keywords{morphing, text-to-audio, environmental sounds}

\section{Introduction} 
Sound morphing refers to the process of gradually transforming one sound into another to generate novel sounds\footnote{\smaller{Audio morphs that have inspired us in our work, e.g. between a baby crying to a trumpet/piano - \href{https://www.cerlsoundgroup.org/Kelly/soundmorphing.html}{https://www.cerlsoundgroup.org/Kelly/soundmorphing.html}}} and hybrid timbres~\cite{kellymorphing,caetano2011morphing,kazazis2016sound}. Such techniques find applications in generating innovative musical compositions and novel sound effects in movies~\cite{kellymorphing}. Recently, diffusion-based~\cite{ho2020denoising} text-to-audio (TTA) models have exhibited remarkable capabilities in generating a wide range of environmental sounds using guidance from text prompts~\cite{liu2023audioldm,liu2023audioldm2, tango_diffusion}. However, their capabilities for gradually or smoothly morphing two sounds are relatively unexplored. 

Most existing systems for morphing are limited to pitched sounds~\cite{kazazis2016sound, caetano2019morphing, caetano2011morphing} or vocal sounds~\cite{slaney1996automatic,ezzat2005morphing}. Such methods use signal processing techniques to extract features such as the coefficients of a source-filter model representation of the two sounds \cite{slaney1996automatic}, or the harmonic components of the sounds~\cite{kazazis2016sound}, to interpolate between them to generate morphs. Although such methods perform well for pitched instruments and voiced utterances, their applicability to inharmonic and noisy environmental sound effects is limited~\cite{gupta2023towards}. 

% Furthermore, directly interpolating between the extracted features may not produce perceptually linear morphs between the sounds~\cite{caetano2011morphing}.

Previously, conditionally trained deep neural networks such as GANs~\cite{engel2017neural,wyse2022sound} have successfully demonstrated their ability to generate morphed instrument sounds while interpolating on its pitch and instrument type in a fine-grained way. Similarly,~\cite{gupta2023towards} show that such GANs can be applied to generate morphs for inharmonic audio textures with specially designed labels. However, such models must be trained or fine-tuned on a small, targeted range of sounds, which limits their applicability to the diverse range of inharmonic environmental sound effects generated by TTA models. Further, such models are unable to provide granular and continuous control over interpolations between the semantics expressed in disparate text prompts to generate morphs. % is difficult.

In this paper, we introduce MorphFader, an interactive technique that utilizes TTA models to morph sounds generated by two different text prompts. In the image domain, Hertz et al.~\cite{hertz2023prompttoprompt} leverage the attention layers within the diffusion process to perform semantic edits to individual images. 
Similarly, we leverage the cross-attention layers in the diffusion process to develop a novel technique for interactive sound morphing. By granularly manipulating the cross-attention components using simple, linear, fader-like controls, we can generate smooth morphs between sounds generated by different text prompts. We evaluate our method objectively using text-audio similarity metrics and subjectively by conducting perceptual listening tests.
% Similarly, we leverage the cross-attention layers in the diffusion process to develop a technique for interactive sound morphing and editing by granularly manipulating the cross-attention components using simple, linear, fader-like controls. We technically evaluate our method objectively using text-audio similarity metrics and subjectively by 
% %perform preliminary user evaluation by 
% conducting perceptual listening tests.

Techniques for audio morphing can be broadly categorized into two - (1) \textit{dynamic morphing}~\cite{kazazis2016sound}, where the source sound gets continuously transformed to the target sound over some time \textit{t}, and (2) \textit{repetitive morphing}~\cite{sethares2015kernel} (also called as \textit{stationary}~\cite{kazazis2016sound}, or \textit{cyclostationary}~\cite{slaney1996automatic} or \textit{static}~\cite{sethares2015kernel} morphing), where a series of intermediate sound morphs are generated, with each progressively containing more features of the target sound and fewer of the source sound. Our work adopts the repetitive morphing paradigm to morph sounds generated by two text prompts. This helps us generate novel intermediate hybrid sounds and timbres that, at times, can generate fantastical sounds at each morph step. 

Our method can operate on any pre-trained TTA models without requiring extra training procedures or fine-tuning. In summary, our contributions include: (a) A novel interactive technique to smoothly morph sounds generated by text prompts and semantically emphasize or ``weight'' certain word descriptors while morphing using pre-trained TTA diffusion models, (b) a systematic comparison of our method with the existing methods through a set of objective and subjective metrics, (c) our code for intercepting and interpolating cross-attention matrices for TTA models. Audio morphs generated using our method can be auditioned on our webpage~\footnote{\smaller{\url{https://pkamath2.github.io/audio-morphing-with-text/}}}.

\begin{figure*}[t]
\vspace{-0.5cm}
\centering
 \includegraphics[width=0.85\textwidth]{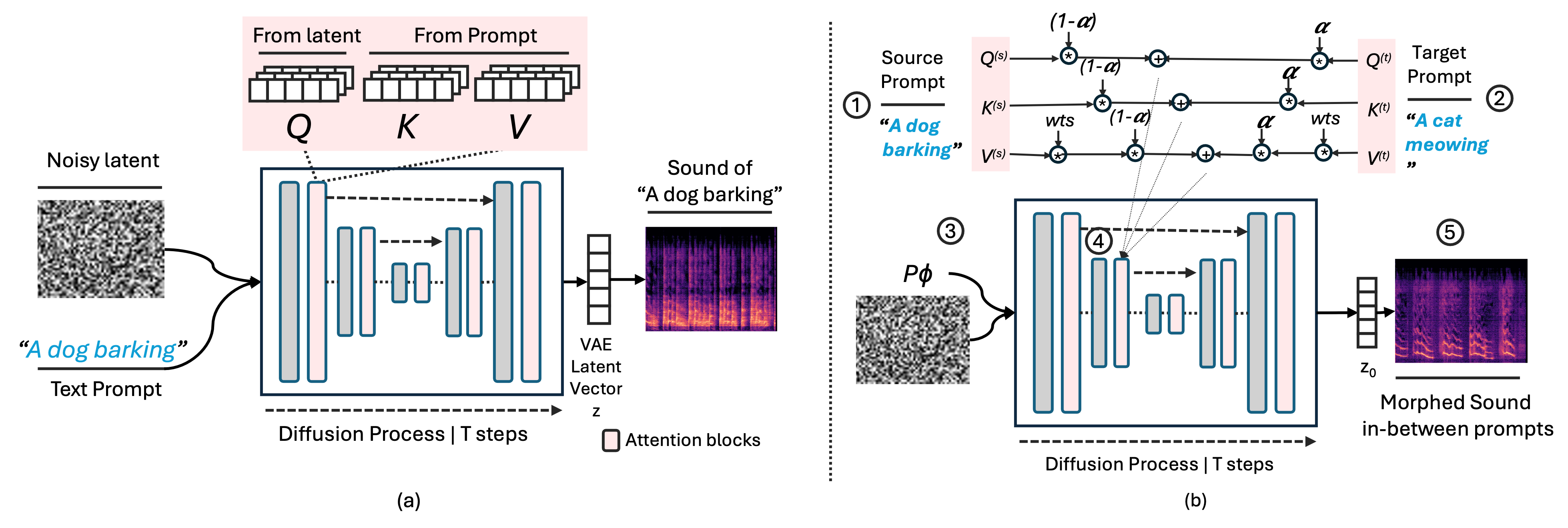}
 \caption{Schematic outlining (a) the diffusion process and (b) our method.}\label{fig:diffusion_arch}
 \vspace{-0.5cm}
\end{figure*}

\section{Background}

At the core of our method is a pre-trained text-to-audio (TTA) latent diffusion model (LDM)~\cite{rombach2022high}. Diffusion models~\cite{ho2020denoising} for audio learn to denoise a spectrogram through a series of steps to generate high-quality sounds. The noise estimates during the denoising process are estimated using a series of U-Nets~\cite{ronneberger2015u}. Text embeddings are injected into the backward denoising process during training to control generation. While diffusion models generally work directly on the spectrogram representations, LDMs, on the other hand, work towards denoising the latent vector representations of a pre-trained Variational Autoencoder (VAE)~\cite{kingma2019introduction}. 

In Figure~\ref{fig:diffusion_arch} (a), we show the schematic of the backward denoising U-Net for one step of the LDM-based diffusion process during inference. The diffusion process accepts a randomly sampled noise vector and a text prompt. Diffusion occurs iteratively in $T$ steps to generate the denoised latent vector $\mathbf{z}$. This latent vector is decoded to a spectrogram using the VAEs decoder network. The spectrogram is converted to an audio waveform using a vocoder~\cite{kong2020hifi}. Note that the details of the diffusion process which we do not modify in our method - such as the pre-trained VAE's encoder and decoder, the diffusion forward process, the vocoder, and the text encoding process - have been skipped in the figure and this paper for brevity. 

In each step of the denoising U-Net are a series of attention layers(shown in pink in Figure~\ref{fig:diffusion_arch}). More specifically, these are cross-attention~\cite{vaswani2017attention} layers, where each word in the text prompt ``attends to" or affects a specific semantic of the generated sound. For instance, a text prompt ``a dog is barking" differs from the prompt ``a dog is barking with reverb" in that the latter also pays ``attention" to the part of the spectrogram that adds reverb to the generated sound. TTA models use cross-attention layers to inject the text prompts into the generative process. More formally, the components of an attention layer are called query $\mathbf{Q}$,  key $\mathbf{K}$, and value  $\mathbf{V}$. Cross-attention is formalized as - 
\vspace{-0.3cm}
\begin{equation}\label{eq:attention}
Cross\textrm{-}Attention( \mathbf{Q},  \mathbf{K},  \mathbf{V}) = \underbrace{\overbrace{Softmax\left( \frac{\mathbf{Q} \mathbf{K}^T}{\sqrt{d}}\right)}^{attention\:map} \: \mathbf{V}}_{cross-attention\:matrix}
\end{equation}

% transformed and projected  ==> embedded
where matrix $\mathbf{Q}$ is the embedded noise vector, and matrices $\mathbf{K}$ and $\mathbf{V}$ are embedded vectors of the text prompt (all shown in Figure~\ref{fig:diffusion_arch} (a)). And ${d}$ is the dimension of the dot product. The $Softmax$ output of the dot product between $\mathbf{Q}$ and $\mathbf{K}$ is referred to as an \textit{attention map}, and the dot product of the attention map and $\mathbf{V}$ is referred to as the \textit{cross-attention matrix}. This cross-attention matrix contains the semantic information from the text prompt and is used to update the spectrogram through the diffusion process. In our work, we focus on manipulating the components of the attention matrices, namely $\mathbf{Q}$, $\mathbf{K}$, and $\mathbf{V}$, for generating morphs and semantically weight (or scale the emphasis of) words in prompts during morphing.

\section{Proposed Framework: MorphFader}
The intuition behind our method is that the components of the cross-attention matrices carry information concerning the semantic similarity between the text prompt and the generated sounds. By ``weighting'' or scaling these components, we can semantically emphasize the presence of a descriptor in the generated sound. Similarly, we can generate perceptually plausible intermediate sound morphs by continuously interpolating between the cross-attention components of two prompts. 

Our method and algorithm are outlined in Figure~\ref{fig:diffusion_arch} (b) and  Equations~\ref{eq:diffusion_process},~\ref{eq:morph_diffusion_process_compinterp},~\ref{eq:morph_diffusion_process_interp}. Say we want to generate a morph between two text prompts - a \textit{source} prompt such as ``A dog barking" and a \textit{target} prompt such as ``A cat meowing." We first run the diffusion process for both prompts separately, and \textit{intercept} and store the $\mathbf{Q}$, $\mathbf{K}$, and $\mathbf{V}$ matrices for each prompt at each time step and each layer in the U-Net, as shown in Eq.~\ref{eq:diffusion_process}.
\begin{equation}\label{eq:diffusion_process}
\mathbf{z_{t-1}}, \mathbf{Q}_t, \mathbf{K}_t, \mathbf{V}_t \gets \mathbf{DM(\mathbf{z_{t}}, \mathcal{P},} t, s)\;
\end{equation}

\noindent where $\mathbf{z_{t}}$ is the noise vector and $\mathbf{Q}_t, \mathbf{K}_t,$ and $ \mathbf{V}_t$ are the attention components at time step $t$ for a particular layer. The layer subscripts are skipped in the equations for brevity. $\mathcal{P}$ and $s$ are the text prompt embedding and random seed, respectively, and $\mathbf{DM}$ is the denoising diffusion step. We then interpolate these matrices between source and target prompts to generate the attention components for the morphed sound. As shown in Eq \ref{eq:morph_diffusion_process_compinterp}, we can interactively control the level of morph or interpolation using a scalar value $\alpha$, where $0<\alpha<1$. 
\vspace{-0.1cm}
\begin{equation}\label{eq:morph_diffusion_process_compinterp}
\begin{gathered}
\mathbf{Q}_t^\mathit{(morph)} \gets \alpha \times\mathbf{Q}_t^\mathit{(\tau)} +(1 - \alpha) \times \mathbf{Q}_t^\mathit{(s)}\\ 
\mathbf{K}_t^\mathit{(morph)} \gets \alpha \times\mathbf{K}_t^\mathit{(\tau)} +(1 - \alpha) \times \mathbf{K}_t^\mathit{(s)}\\
\mathbf{V}_t^\mathit{(morph)} \gets \alpha \times\mathbf{V}_t^\mathit{(\tau)} +(1 - \alpha) \times \mathbf{V}_t^\mathit{(s)}\\
% \mathbf{z^{\mathit{(morph)}}_{t-1}} \gets 
% \mathbf{DM(\mathbf{z_{t}}, \mathcal{P^{\phi}},} t, s)\{\mathbf{Q}_t^\mathit{(morph)}, \mathbf{K}_t^\mathit{(morph)}, \mathbf{V}_t^\mathit{(morph)}\}\\
\end{gathered}
\end{equation}

\noindent where superscripts $\mathit{s}$, $\mathit{\tau}$, and $\mathit{morph}$ indicate source, target, and morphed components respectively. As shown in Eq.~\ref{eq:morph_diffusion_process_interp}, we inject these morphed attention components into the diffusion process while generating the morphs.
\vspace{-0.1cm}
\begin{equation}\label{eq:morph_diffusion_process_interp}
\begin{gathered}
\mathbf{z^{\mathit{(morph)}}_{t-1}} \gets 
\mathbf{DM(\mathbf{z_{t}}, \mathcal{P^{\phi}},} t, s)\{\mathbf{Q}_t^\mathit{(morph)}, \mathbf{K}_t^\mathit{(morph)}, \mathbf{V}_t^\mathit{(morph)}\}\\
\end{gathered}
\end{equation}

\noindent where $\mathcal{P^{\phi}}$ indicates an unconditional or empty string prompt embedding. Note that the attention components generated using $\mathcal{P^{\phi}}$ are ignored, and the injected morphed components are used instead. As $\alpha$ changes from $0$ to $1$, the morph slowly changes from the source to the target sound. The final morphed latent vector $\mathit{z_0}$ generated at the end of the diffusion step $T$ is decoded using the VAE decoding process to generate the morphed sound. Note that the above process is run for each attention layer within the U-Net. The full algorithm can be viewed on our webpage. 

In the morphing process above, the matrix $\mathbf{V}$ can be further word-weighted to increase or decrease the emphasis of the verb ``bark" or the ``meow" in the resulting sound. 

\begin{equation}\label{eq:weighting_v}
\overline{\mathbf{V}} = \mathbf{wts}\times\mathbf{V}
\end{equation}

\noindent where $\mathbf{V}$ is the original value matrix, $\mathbf{wts}$ is the weight vector, and $\overline{\mathbf{V}}$ is the resulting semantically weighted value matrix. Our weighting approach $\mathbf{V}$ achieves similar goals to the semantic editing method for images outlined in~\cite{hertz2023prompttoprompt}. In~\cite{hertz2023prompttoprompt}, authors propose to weight the full attention map for performing edits. Instead, empirically, we find it more computationally efficient to intercept, interpolate, and inject individually weighted $\mathbf{V}$ components than the full dot-product attention map through each layer and per step of the diffusion process while morphing or word-weighting sounds. By interpolating between the attention components of the two prompts in this way, we can generate fantastical animal vocalizations, such as a morph between a dog's ``bark'' (source) and a cat's ``meow'' (target). 

\section{Experimental Setup}
\noindent \textbf{Implementation Details:}
We implement our method over a pre-trained text-to-audio model AudioLDM~\cite{liu2023audioldm}. Specifically, we use the \textit{``audioldm\_16k\_crossattn\_t5''} model, which uses cross attention and is finetuned on FLAN-T5~\cite{chung2022flant5} embeddings. Although we demonstrate the effectiveness of our method using AudioLDM, our algorithm can easily integrate with any LDM that uses cross-attention (such as TANGO~\cite{tango_diffusion} or Stable Audio~\cite{evans2024fast}). We run our experiments on an RTX 2080 Ti 11GB GPU. All samples were generated using a constant random seed and by running diffusion for $T=20$ steps. A demo video, examples, and codebase can be found on our webpage. All sounds generated in this paper are $10$ seconds long.

\noindent \textbf{Datasets:}
We sourced text prompts from a dataset called \textit{AudioPairBank}~\cite{sager2018audiopairbank} to evaluate our morphing technique. The AudioPairBank dataset contains over $1123$ adjective-noun and verb-noun text-based acoustic concept pairs mined from databases such as FreeSound (FS). It associates an adjective or a verb with nouns to create concept pairs such as a ``barking dog," etc. 

\noindent \textbf{Evaluation Metrics:} Following Liu et al.~\cite{liu2023audioldm}, we measure the quality of our morphs using audio quality metrics such as \textbf{\textit{Fréchet Audio Distance (FAD)}}, \textbf{\textit{Fréchet Distance (FD)}}, and \textbf{\textit{Inception Score (IS)}}. FAD~\cite{kilgour2019frechet} is the distance between the distributions of the embeddings of real and synthesized audio data extracted from a pre-trained VGGish model. FD is similar to FAD but uses state-of-the-art audio classifier PANN~\cite{kong2020panns} for embeddings. Lower values for both are better. IS evaluates the quality and diversity of audio using the PANN classifier. Higher IS values are better. We compute two sets of FAD and FD metrics: (1) FAD-AudioSet and FD-AudioSet using $5000$ randomly sampled audio files from the AudioSet~\cite{gemmeke2017audio} evaluation dataset as a reference, and (2) FAD and FD using $200$ samples of source and target sounds (generated from AudioLDM) as reference, that were used for generating the morphs.

Morphing is a creative task and is typically assessed based on the subjective aesthetics of the sound. Caetano et al.~\cite{caetano2012formal} suggest measures such as \textbf{\textit{`smoothness'}} to objectively evaluate morphs. They define the `smoothness' of a morph as the ability of the method to morph the sound from source to target linearly. So, we use perceptual linearity metrics derived from text-audio similarity scores based on CLAP~\cite{elizalde2023clap}. We measure the linearity of change in the score w.r.t the morph interpolation step $\alpha$ and compute it using the Pearson correlation coefficient ($\rho$)(as in~\cite{gupta2022parameter}). Higher is better. Finally, we use \textbf{\textit{Mean Opinion Scores} (MOS)} for evaluations using listening tests.

\noindent\textbf{Baseline Selection:}
While selecting baselines for our experiments, we found that existing state-of-the-art toolkits, such as sound morphing toolbox~\cite{caetano2019morphing}, fail for non-pitched sounds. Further, other deep learning methods, such as in~\cite{gupta2023towards}, generate morphs for only a small targeted range of sounds, such as wind or water. To the best of our knowledge, there is currently a lack of methods to morph inharmonic general-purpose environmental sounds, such as those generated using TTA models. Thus, for our baseline comparison, we selected two handcrafted methods - (1) linearly interpolating or mixing source and target raw audio waveforms and (2) morphing using engineered text prompts. For this, we used engineered prompts such as ``A morph between \texttt<Sound A\texttt> and \texttt<Sound B\texttt> where the level of \texttt<Sound A\texttt> is at \texttt<X\texttt>\% and level of \texttt<Sound B\texttt> is at \texttt<(100-X)\texttt>\%'' to generate morphs. X is percentage interpolation level ($\alpha * 100$).

% When choosing baselines for our experiments, we found that existing state-of-the-art toolkits, such as sound morphing toolbox~\cite{caetano2019morphing}, and deep learning methods, such as in~\cite{gupta2023towards,wyse2022sound}, were not suitable for morphing inharmonic general-purpose environmental sounds. As a result, we selected two handcrafted methods for comparison: interpolating or mixing using raw audio waveforms, and morphing using engineered text prompts.
\vspace{-0.2cm}
\subsection{Experiments \& Results}

\subsubsection{Ablation Studies}
\vspace{-0.2cm}
We first conduct ablation studies by systematically ablating each individual $\mathbf{Q}$, $\mathbf{K}$, $\mathbf{V}$ component during morphing to understand its effect on the generated morph in Equation~\ref{eq:morph_diffusion_process_interp}. We randomly sampled $100$ source-target prompt pairs from AudioPairBank to generate the sounds for this experiment. We generated sounds and intercepted attention components for the individual prompts. We then interpolated the attention components granularly using our method, using $\alpha$ in steps of $0.1$ between the range $[0,1]$ to generate $11$ linearly morphed sounds for each source-target prompt pair. This experiment generated $1100$ morphed sounds for evaluation.

Table~\ref{tab:ch4-ablation} shows the FAS-AudioSet, FD-AudioSet, FAD, FID, IS, and Smoothness scores for this experiment. $(\downarrow)$ indicates that lower values are better. We find that using $\mathbf{Q, K, V}$ and $\mathbf{K, V}$ outperforms other attention component combinations. We use the best performing $\mathbf{Q, K, V}$ for all experiments in the remainder of the paper.

\begin{table}[t]
    \caption{Ablation Studies}
    \label{tab:ch4-ablation}
    \vspace{-0.2cm}
    \scalebox{0.85}{
    \centering
    \begin{tabular}{|c|c|c|c|c|c|c|}
    \hline
    &FAD ($\downarrow$)&FD ($\downarrow$)& FAD ($\downarrow$)& FD ($\downarrow$)&IS($\uparrow$)&Smooth-($\uparrow$)\\
    &AudioSet&AudioSet&&&&ness\\
    \hline
    $\mathbf{Q}$,$\mathbf{K}$,$\mathbf{V}$&$\boldsymbol{10.81}$&56.68&$\boldsymbol{0.25}$&$\boldsymbol{5.14}$&$\boldsymbol{5.98}$&$\boldsymbol{0.61}$\\
    \hline
    $\mathbf{K}$,$\mathbf{V}$&10.82&$\boldsymbol{56.61}$&0.26&$\boldsymbol{5.14}$&5.96&0.60\\
    \hline
    $\mathbf{Q}$,$\mathbf{K}$&17.53&94.71&7.48&50.79&1.80&0.30\\
    \hline
    $\mathbf{Q}$,$\mathbf{V}$& 12.73&81.72&4.87&42.72&2.54&0.41\\
    \hline
    $\mathbf{Q}$ only& 17.54&94.71&7.47&50.80&1.80&0.31\\
    \hline
    $\mathbf{K}$ only& 27.09&134.35&14.74&96.78&1.00&0.30\\
    \hline
    $\mathbf{V}$ only&12.73&81.72&4.87&42.72&2.54&0.40\\
    \hline
    \end{tabular}
    }
    \vspace{-0.5cm}
\end{table}

\vspace{-0.2cm}
\subsubsection{Baseline Comparison}
\vspace{-0.2cm}
\begin{table*}
% \parbox{.45\linewidth}{
    \caption{Morphing Baseline Comparison}
    \label{tab:ch4-baseline}
    \vspace{-0.2cm}
    \centering
    % \scalebox{0.8}{
    \begin{tabular}{|c|c|c|c|c|c|c|c|}
    \hline
    &FAD ($\downarrow$)&FD ($\downarrow$)& FAD ($\downarrow$)& FD ($\downarrow$)&IS($\uparrow$)&Smoo- ($\uparrow$)& MOS ($\uparrow$)\\
    &AudioSet&AudioSet&&&&thness&\\
    \hline
    Ours&10.81&56.68&$\boldsymbol{0.25}$&$\boldsymbol{5.14}$&$\boldsymbol{5.98}$&$\boldsymbol{0.61\pm\scriptscriptstyle{0.03}}^*$&$\boldsymbol{50.49}\pm\scriptscriptstyle{1.66}$ \\
    \hline
    Wavform Mixing&$\boldsymbol{9.13}$&$\boldsymbol{52.19}$&0.92&12.88&5.34&$\boldsymbol{0.61\pm\scriptscriptstyle{0.07}}^*$&$29.50\pm\scriptscriptstyle{1.91}$\\
    \hline
    Prompting& $11.73$&$67.10$&$1.53$&$18.21$&$5.20$&$0.34\pm\scriptscriptstyle{0.03}$&$45.26\pm\scriptscriptstyle{1.90}$\\
    \hline
    \end{tabular}
    \vspace{-0.4cm}
\end{table*}

% \begin{figure}[t]
% \centering
%  \includegraphics[width=0.85\linewidth]{resources/together_morph_plot.png}
%  \caption{Plot for Text-Audio Similarity Scores while morphing.}\label{fig:together_morph_plot}
% \end{figure}

To objectively compare our morphing method (Equation~\ref{eq:morph_diffusion_process_interp}) with the selected baselines, we randomly selected $100$ source-target prompt pairs. We generated $1100$ linearly morphed samples using our method following the same procedure outlined in ablation studies. For generating sounds using waveform mixing baseline, we granularly interpolated the source and target prompted raw-audio waveforms to generate $1100$ mixed sounds. For morphs generated using engineered text prompts baseline, we crafted $1100$ prompts by modifying the level values based on $\alpha$ in the prompt to generate interpolated morphs between the source and target.

Table~\ref{tab:ch4-baseline} shows our method's results compared with the selected baselines. $(\downarrow)$ indicates lower scores are better. Our method can generate better-quality sounds in terms of FAD, FD, and IS compared to the baselines. The mixes generated interpolating raw-audio waveforms demonstrate better FAD-AudioSet and FD-AudioSet scores than our method. Interestingly, our method and waveform mixing perform equally well when evaluated on the smoothness metric. A two-way t-test indicates there were no significant differences between the two smoothness scores (`\textbf{*}' in the table, $p>0.05$). However, by qualitatively listening and comparing the morphs generated by the two methods, we find that the sounds generated by our method generate perceptually novel sounding elements and are not simply an additive mix of the source and the target. We encourage our readers to audition the sounds for comparisons on our webpage.

\noindent\textbf{Listening Tests:}
We conduct listening tests by recruiting $N=18$ participants to subjectively analyze our method's effectiveness in generating morphs compared with the two baselines and report mean opinion scores (MOS). %To generate sounds for our listening test,
We randomly sampled $20$ source-target prompt pairs from the AudioPairsBank, and generated morphs (at $\alpha=0.5$) using our method and the two baselines. The test was administered online and can be viewed on our webpage. The participants were asked to complete the test in a single sitting and requested to use noise-cancellation headphones during the test. 
% They were also asked to undertake the test in a quiet environment.

First, we instructed our participants to audition a popular example of a good morph of a baby crying and piano\footnote{\smaller{We chose the morph of a baby crying to piano from \href{https://www.cerlsoundgroup.org/Kelly/soundmorphing.html}{cerlsoundgroup.org}}}. We provided them with an instruction to evaluate the morphs: ``During the evaluation, ask yourself - \textit{`how would I imagine a baby crying to the tune of a piano?'} and score the option closest to it higher than the rest''. For each listening trial, we asked participants to listen to source and target sounds and score each of the three presented morphed sound examples for their perceptual plausibility on a scale from $[0-100]$. 

Table~\ref{tab:ch4-baseline} shows the MOS from our listening test. $(\uparrow)$ indicates higher values are better. Participants rated morphs generated using our method as perceptually better as compared to mixes generated using raw-audio waveforms $(t(17)=11.52, p<0.05)$ as well as engineered prompts $(t(17)=2.70, p<0.05)$. The subjective and objective evaluation results show that MorphFader is able to effectively generate perceptually plausible morphs. We encourage our readers to audition the sounds on our webpage to gauge the effectiveness of our method in comparison with the two baselines.
\vspace{-0.2cm}
\subsubsection{Evaluating Word Types}
\vspace{-0.2cm}
\begin{table}[t]    \caption{Analyzing Word Types}
    \label{tab:ch4-wordtypeeval}
    \vspace{-0.2cm}
    \scalebox{0.8}{
    \centering
    \begin{tabular}{|c|c|c|c|c|}
    \hline
    &\multicolumn{2}{|c|}{Word-weighting}&\multicolumn{2}{|c|}{Morphing}\\
    \hline
    &Smooth-& MOS ($\uparrow$)&Smooth-($\uparrow$)& MOS ($\uparrow$)\\
    &ness($\uparrow$)&&ness($\uparrow$)&\\
    \hline
    Adjective-&${0.23\pm\scriptscriptstyle{0.03}}$&$0.55\pm\scriptscriptstyle{0.04}$&$0.46\pm\scriptscriptstyle{0.18}^*$&$0.55\pm\scriptscriptstyle{0.10}^*$\\
    based prompts&&&&\\
    \hline
    Verb&$\mathbf{0.56\pm\scriptscriptstyle{0.06}}$&$\mathbf{0.68\pm\scriptscriptstyle{0.06}}$&$0.61\pm\scriptscriptstyle{0.15}^*$&$0.69\pm\scriptscriptstyle{0.07}^*$\\
    based prompts&&&&\\
    \hline
    \end{tabular}
    }
    \vspace{-0.5cm}
\end{table}

In this experiment, we study the effect of morphing (Equation~\ref{eq:morph_diffusion_process_interp}) and semantic word-weighting (Equation~\ref{eq:weighting_v}) adjectives in prompts compared to verbs. To analyze semantic word-weighting sounds, we randomly sampled $100$ adjective-based and $100$ verb-based prompts from the AudioPairBank. We linearly modified the weights on the adjective or verb descriptors from $[-2,3]$ in steps of $1$ to generate overall $600$ linearly word-weighted sounds. Similarly, we sampled $100$ adjective-based and verb-based source-target prompt pairs and interpolated $\alpha$ in steps of $0.1$ to generate $1100$ morphed sounds to perform this evaluation.

\begin{figure}[t]
\centering
 \includegraphics[width=0.7\linewidth]{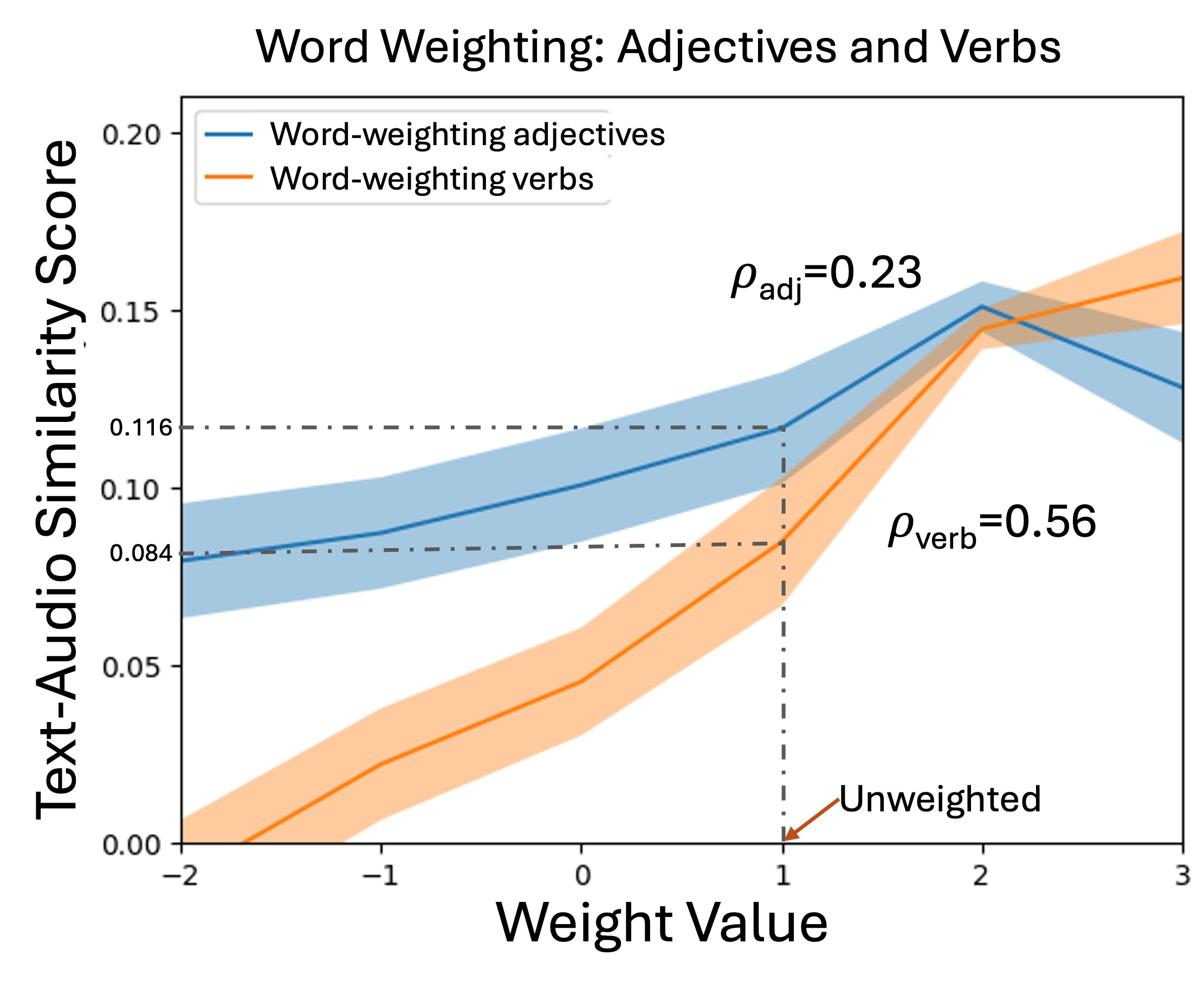}
 \vspace{-0.3cm}
 \caption{Plot for Text-Audio Similarity Scores for word-weighting.}\label{fig:word_weight_plot}
 \vspace{-0.5cm}
\end{figure}

Table~\ref{tab:ch4-wordtypeeval} shows scores for smoothness (or linearity) of word-weighting and morphing. We observe that word-weighting verbs in the text prompts were significantly smoother ($\rho=0.56$) than weighting adjectives ($\rho=0.23$). Figure~\ref{fig:word_weight_plot} visualizes the smoothness of interpolation at each step between $[-2, 3]$ for verb and adjective descriptors. The dotted line shows the similarity score at word-weight $\mathbf{=1}$, i.e., unweighted generation. Shaded regions show standard error of means computed by bootstrapping.
% over 100 iterations.

Table~\ref{tab:ch4-wordtypeeval} also shows scores for the smoothness of morphing when using prompts with adjectives ($\rho = 0.46$) and verbs ($\rho = 0.61$). There were no significant differences between the smoothness scores for both prompts $(p>0.05)$. This indicates that our method can morph both adjective- and verb-based prompts equally well.

\noindent\textbf{Listening Tests:} We conducted a listening test by recruiting $N=17$ participants to subjectively analyze the effect of word types. For word weighting, we randomly selected $5$ adjective-based and verb-based prompts each and adjusted the word weights by $-1$ and $+1$. We asked the participants to evaluate generated sounds for quality of semantic edit w.r.t to the reference unweighted sound. For the morphing evaluation, we randomly selected $4$ source-target pairs of adjective-based and verb-based prompts each and generated a morphed sound with $\alpha=0.5$. The participants were asked to evaluate the plausibility of the morphed sound. Each participant attempted $20$ word-weighting sound trials and $8$ morphing trials.

Table~\ref{tab:ch4-wordtypeeval} shows the MOS scores for this experiment. A two-sampled t-test for word-weighting revealed our listeners could better evaluate semantic changes to verb-based descriptors than adjectives ($(t(16)=-2.39$,$ p<0.05)$). The t-test for morphing, however, revealed no significant differences, i.e., our listeners evaluated morphs between adjectives and verbs as equally plausible or ``in-between'' the source and target sounds$(p>0.05)$.

This result has implications when designing controls using adjective- or verb-based text prompts for audio generation. Verbs are less subjective and more neutral, making them easier for listeners to identify in sounds~\cite{sager2018audiopairbank}. For example, while annotating (eg.~captions, tags) audio datasets, there is less subjective debate about the presence of a barking sound (verb) than the size or type of the dog (adjective).
%This subjectivity may be prevalent in the captions and tags of large audio datasets.
Therefore, controls in audio editing tools that modify text-based semantics using verbs would be more effective than those using adjectives.
\vspace{-0.3cm}
\section{Conclusion}
\vspace{-0.25cm}
This paper introduced an interactive technique for morphing sounds generated by pre-trained text-to-audio (TTA) models. Our method intercepts and interpolates between the attention components from cross-attention layers within the diffusion process to generate morphs. With no additional training or fine-tuning, our method generates smooth sound edits and perceptually plausible morphs between sounds generated by different text prompts. We validated our approach objectively and subjectively through listening tests.

% \balance{}

\bibliographystyle{IEEEbib}
\bibliography{paper.bib}
\label{sec:refs}

\end{document}